\documentstyle[12pt,epsf]{article}
\newcommand{\lsim}{ \raisebox{-.5ex}{\mbox{$\,\stackrel{<}{\sim}$\,}} }

\begin{document}

\vspace{0.7cm}

 \begin{center}
{\large\bf Critical perturbation in standard map:\newline
A better approximation}\\

\vspace{0.5cm}

         Boris Chirikov\footnote{Email: chirikov@inp.nsk.su}\\ [5mm]
              {Budker Institute of Nuclear Physics\\
               630090 Novosibirsk, Russia}
\end{center}

\vspace{1cm}

\begin{abstract}
Direct computation of the transition time between neighbor resonances
in the standard map, as a function of the perturbation parameter $K$,
allows for improving the accuracy of the critical perturbation value up to
$K_{cr}-K_g < 2.5\times 10^{-4}$ that is by a factor of about 50 as compared to
the previous result due to MacKay and Percival.

\end{abstract}

\vspace{1cm}

As is well known by now a typical structure of the phase space
of a few--freedom nonlinear dynamical system is characterized 
by a very complicated admixture of both chaotic as well as
regular (integrable) components of motion (the so--called divided
phase space, see, e.g., \cite{chirikov,lib,chirev}). Statistical properties
of such a motion are very intricate and unusual. One of the most
interesting (and important for many applications) problem is the
conditions for transition from a local (restricted to relatively
small regions in phase space) to the global chaos covering the whole
available phase space. The most studied model of such a transition
is described by the so--called (canonical) standard map 
(for history of this model see \cite{chizetp}):
\begin{equation}
\overline {y} = y - \frac{K}{2 \pi}\cdot \sin (2\pi x) ~~,~~ 
\overline {x}= x + {\overline y}
\label{stmap}
\end{equation}
where $K$ is the perturbation parameter. In this simple model the transition 
to global chaos corresponds to some exact critical value $K=K_{cr}$.
For $K>K_{cr}$ the motion becomes infinite (in momentum $y$) for some 
initial conditions while for $K\leq K_{cr}$ all the trajectories are
confined within a period of map (\ref{stmap}): $\Delta y=1$.

The first idea how to solve this difficult problem was due to Greene
\cite{green}. First, he was able to solve a much simpler problem
of the critical perturbation $K(r)$ at which a particular invariant 
Kolmogorov - Arnold - Moser (KAM) curve with the rotation number $r$ 
is destroyed. Critical function $K(r)$ is extremely singular with big
dips at everywhere dense set of rational $r$ values (see, e.g., \cite{stark}).
The physical mechanism of this behavior (known since Poincar\'e) 
is explained by 
resonances in the system (\ref{stmap}) as the rotation number is the ratio of
oscillation/perturbation frequencies. Whence, the main Greene's idea:
to find the 'most irrational' $r=r_g$ which would correspond to the
motion 'most far--off' all the resonances. The former is well known
in the number theory: $r_g=[111...]=(\sqrt{5}-1)/2$ where the first
representation is a continued fraction.
This 'golden' curve was found to be critical
at the parameter $K=K_g=0.97163540631...$ \cite{green,mackay}. 
It was conjectured that
for $K>K_g$ all invariant 
curves are destroyed \cite{mackay}, that is $K_{cr}=K_g$.

The 'most--irrational' assumption - as plausible as it is - remains a
hypothesis. The main difficulty is here in that the resonance interaction 
and overlap, destroying invariant curves, depend not only on the resonance
spacings, which are indeed maximal for $r=r_g$, but also on the 
amplitudes of those which are not simply an arithmetical property.
Another argument, based on the analysis of the critical function $K(r)$
\cite{bial,murray}, also does not prove this principal hypothesis.

A different approach to the problem - the so--called converse KAM theory - was
developed
in \cite{perci,inkam}. It relies upon a rigorous criterion for the absence of
any invariant curve in a certain region. Unfortunately, this criterion
can only be checked numerically, and besides it provides the upper bound
$K^{+}_{cr}$ only (the lower bound $K^{-}_{cr}=K_g$). The remaining gap,
or the accuracy of $K_{cr}$:
\begin{equation}
(\Delta K)_{cr}\,=\,K^{+}_{cr}\,-\,K_{cr}
\label{gap}
\end{equation}
can be made arbitrarily small at the expense of computation time $t_C$
which scales as \cite{perci} 
\begin{equation}
t_C\,\propto\,(K^{+}_{cr}\,-\,K_{cr})^{-p}
\label{scale1}
\end{equation}

\begin{figure}[]
\centerline{\epsfxsize=15cm \epsfbox{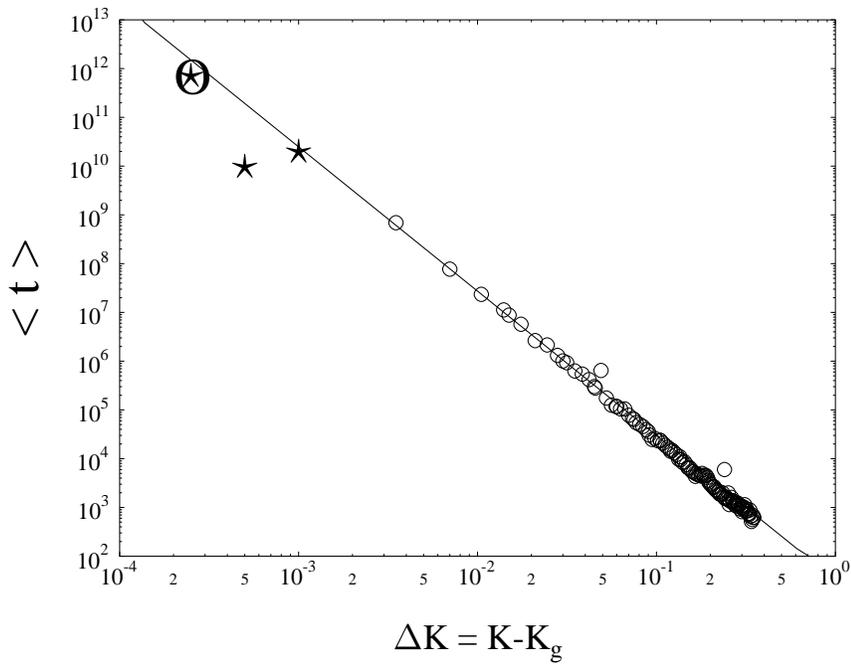}}
\caption{Direct computation of $K_{cr}$ in standard map:
             average transition time through the destroyed critical curve
             vs. supercriticality. Circles show numerical results for
             $N_{tr}=400$; stars represent 3 single--trajectory runs,
             including one with the minimal $\Delta K$ (\ref{abac});
             straight line is relation (\ref{scale2}) with parameters
             (\ref{stac}) fitted from 15 left--most points (circles).
}
\end{figure}

Facing this difficulty, it is natural to recall the first method
for calculating the critical perturbation used in \cite{chirikov}.
The method was based on the direct computation of trajectories for
different $K\to K_g$. The criterion of supercriticality of a particular
$K$ value was very simple: the transition if only a single trajectory 
in one of two neighbor integer resonances ($y_r=0\ mod\ 1$) through the
destroyed critical curve.
With the computers available at that time
the minimal $K=1$ has been reached only which corresponds to the
uncertainty $(\Delta K)_{min}=K_{min}-K_g=0.0284$. This may be compared
to the later result $(\Delta K)_{min}=0.0127$ \cite{perci}.

Remarkably, the dependence of the average transition time on parameter $K$
was found to be similar to scaling (\ref{scale1}):
\begin{equation}
<t>\,=\,\frac{A}{(K\,-\,K_{cr})^{p}}
\label{scale2}
\end{equation}
Fitting three unknown parameters gave: $A=103,\ p=2.55$, and $K_{cr}=0.989$.
The latter result was rather different from the present value
$K_{cr}\approx K_g$, again because of the computation restrictions mentioned
above: $K\geq 1,\ t\leq 10^7$ iterations. 
Nevertheless, the fitting Eq.(\ref{scale2})
provided a less uncertainty $(\Delta K)_f=K_f-K_g=0.0174$ as compared to
the result from the minimal $K$. The same is true for data
from \cite{perci} where $(\Delta K)_f=K_f-K_g=0.00236$. 
The latter value  was
apparently obtained by the direct fitting the relation (\ref{scale1}).
Fitting in log--log scale provides a much better result: 
$(\Delta K)_f=K_f-K_g=-0.000128\pm 0.000288$ that is the remainig uncertainty
reduces down to $0.000288$.

In both
cases the fitted value for the critical perturbation $K_{cr}$ is 
only true up to a
certain confidence probability while the minimal $K$ is an exact result:
$K^+_{cr}=K_{min}$.

In the present paper the studies \cite{chirikov} are continued with much
better computers. The main result is farther considerable increasing of
the accuracy $(\Delta K)_{cr}$.

To reduce the computation expenses, the transition time was calculated for a
number of trajectories $N_{tr}$ started near the unstable fixed point of a 
half--integer resonance ($y_r=1/2\ mod\ 1$), and then run until 
each of them crossed over to a neighbor integer resonance. 

The minimal $K$ value is determined already by the first trajectory escaped
from the half--integer resonance. In this way the minimal uncertainty
\begin{equation}
(\Delta K)_{min}=K_{min}-K_g=0.00025
\label{abac}
\end{equation}
has been achieved with the escape time $t\approx 6.77\times 10^{11}$
itterations which took about 72 hours of CPU time on ALPHA--4100
computer (see Fig.1).

The average transition time was computed from $N_{tr}=400$ trajectories
for each of 100 values of $K$ in the interval: 
$0.0035\leq K-K_g\leq 0.35$. This costed 36 hours of computation.
The results are shown in Fig.1.
In the whole interval of $\Delta K$ the dependence $<t(K)>$ 
is not exactly a power--law.
It becomes so asymptotically for $K\to K_{cr}$ as expected from the theory
\cite{meiss}. For this reason, only few smallest $K$ values of the function
$<t(K)>$ were taken for the final fitting which is also shown in Fig.1 
by the solid line. It is obtained from the fitting 15 left--most points
(just up to the first big fluctuation) in log--log scale, and
corresponds to the following parameters in Eq.(\ref{scale2}):
\begin{equation}
(\Delta K)_f\,=\,0.000125\,\pm 0.000267, \qquad p\,=\,2.959\,\pm 0.0771, 
\quad A\,=\,33\,\pm 8
\label{stac}
\end{equation}
The fitting relative accuracy $rms=0.071$ is close to, but somewhat 
larger than, 
the standard $rms=1/\sqrt{N_{tr}}=0.05$. This is seen from the data
of 3 single trajectories in Fig.1, too. Notice also 2 very big 
deviations for the
average over 400 trajectories which nature remains unclear.
Interestingly, the relative fitting accuracy of the data \cite{perci}
is considerably higher: $rms=0.02$. This would require as many as about
5000 trajectories in the present method. However, it does not mean that
the computation of the procedure in \cite{perci} would be shorter.

The most important parameter in (\ref{stac}) is $(\Delta K)_f$ 
which is zero within
statistical errors. This farther confirms the Greene hypothesis
$K_{cr}=K_g$. The exponent $p$ is also equal to the theoretical value
$p_{th}=3.011722$ \cite{meiss} to the fitting accuracy.
The present value of parameter $A$ is much less than in
\cite{chirikov} because of a different (shorter) transition
between resonances used.
The summary of all results is presented in the Table below.
\vspace{1cm}
{\large
\begin{center}
\begin{tabular}{|c|c|c|}
\multicolumn{3}{l}{Table. Accuracy of $K_{cr}$ in standard map}\\ [3mm] \hline
\multicolumn{1}{|c|}{$(\Delta K)_{min}$} &
\multicolumn{1}{|c|}{$(\Delta K)_{fit}$} &
\multicolumn{1}{|c|}{Reference} \\ 
exact & probable & \\ \hline
$2.84\times 10^{-2}$ & $1.74\times 10^{-2}$ & \cite{chirikov} \\ \hline
$1.27\times 10^{-2}$& $3.36\times 10^{-3}$ & \cite{perci} \\
         & $\pm 1.\times 10^{-3}$ & \\ \hline
         & $-1.28\times 10^{-4}$  & \cite{perci} \\
         & $\pm 2.88\times 10^{-4}$ & our fit \\ \hline
$2.5\times 10^{-4}$ & $1.25\times 10^{-4}$ & present \\
                    & $\pm 2.67\times 10^{-4}$ & paper \\ \hline 
\end{tabular}
\end{center}
}

A serious difficulty in such a numerical approach to the problem 
is the computation accuracy.
This was mentioned also in \cite{perci} but no estimate for the 
computation errors
was given, apparently because of a very complicated numerical procedure.
Even in a much simpler method \cite{chirikov}, accepted in the present study,
the effect of noise turned out to be rather complicated. 
Special numerical experiments were done to clarify the question.
To this end, a random perturbation of amplitude $\nu$ was introduced
in both equations (\ref{stmap}). The results are shown in Fig.2.

\begin{figure}[]
\centerline{\epsfxsize=15cm \epsfbox{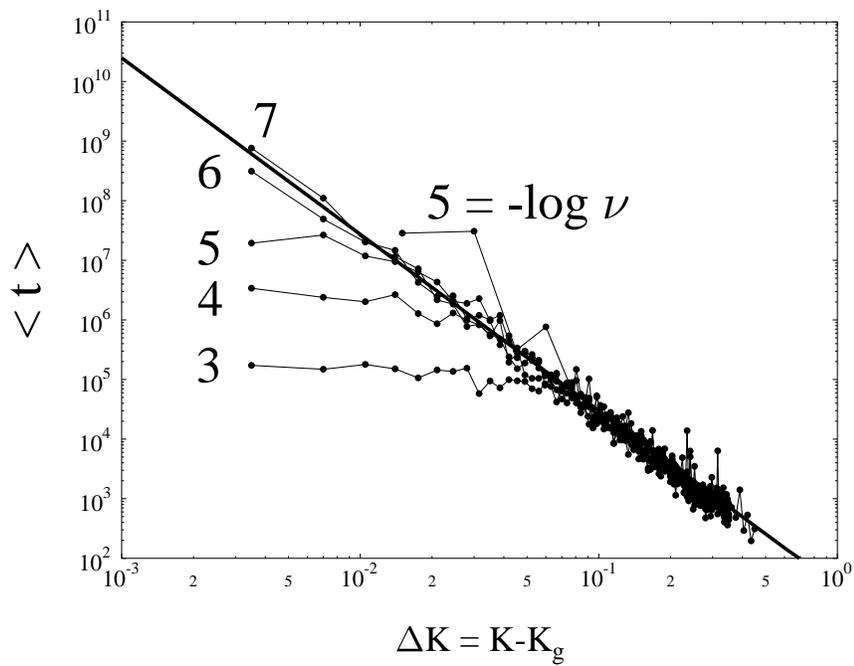}}
\caption{The effect of noise on the supercritical transition time:
             straight line is the fit of the noisefree 
             computation results (cf. Fig.1); points connected by lines
             represent the impact of noise with amplitude $\nu$ 
             computed for $N_{tr}=10$; numbers at lines are -log$(\nu )$
             values (logarithm decimal).
}
\end{figure}

\begin{figure}[]
\centerline{\epsfxsize=15cm \epsfbox{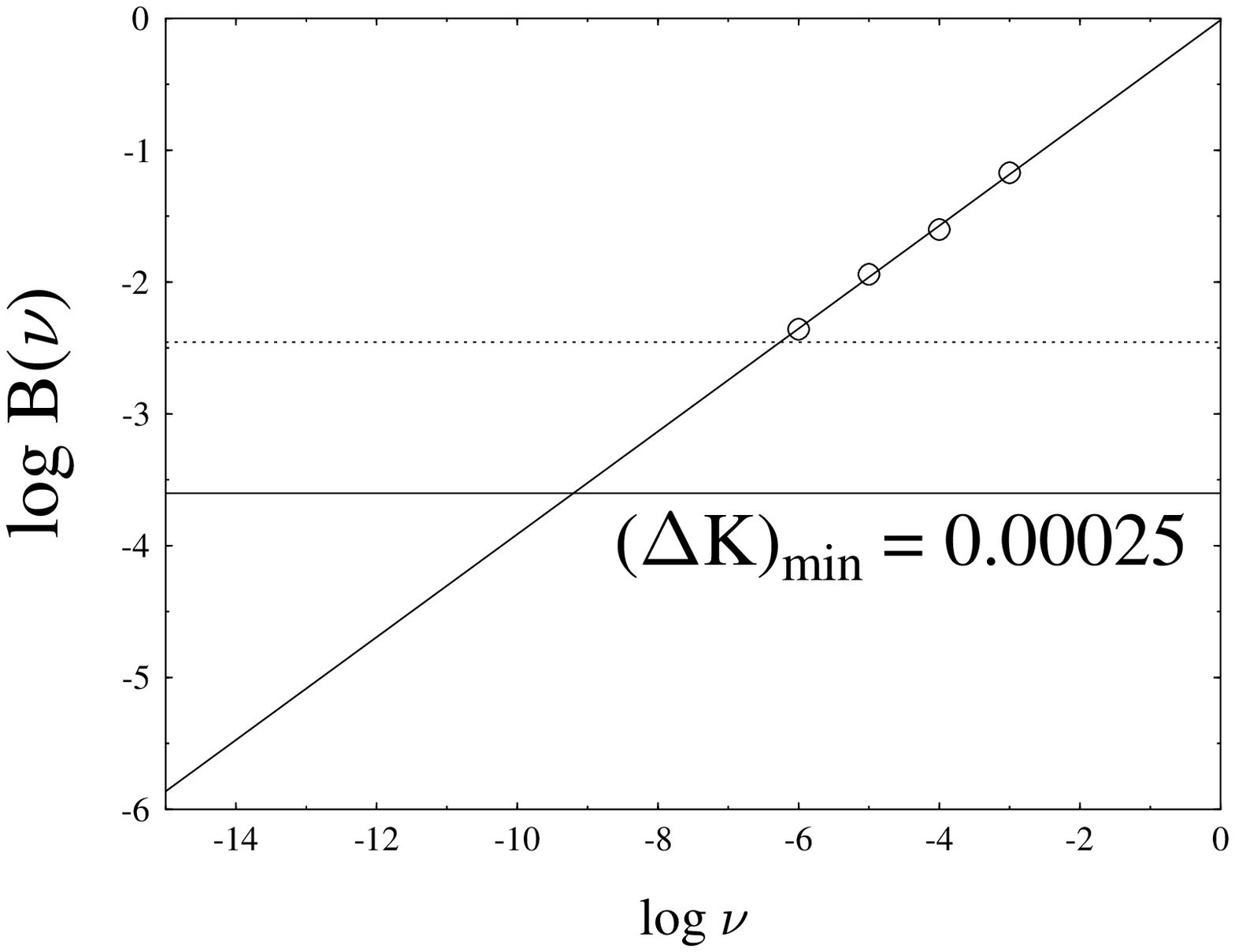}}
\caption{Noise scaling: circles give the empirical crossover values 
             $B(\nu )$
             as a function of noise amplitude $\nu$ connected and extrapolated
             by the straight line (\ref{noise}); the upper horizontal
             dotted line
             shows the minimal $\Delta K$ in computation with noise
             while the lower line indicates $(\Delta K)_{min}$ (\ref{abac})
             achieved in the main double--precision computation (see Fig.1)
             with the accuracy roughly corresponding to 
             log$(\nu )\approx -15$; all logarithms are decimal.
}
\end{figure}

Typically, the transition time becomes less than that without noise, and 
saturates below some critical
noise--dependent value of $K$: $\ \Delta K\lsim B(\nu )$.
However, in some cases the average transition time considerably grows,
as an example in Fig.2 demonstrates, apparently due to a sharp increase
of the fluctuations near the crossover from normal (noisefree) dependence
of $<t(\Delta K)>$ to the saturation. In turn, these fluctuations are
apparently explained by the noise--induced diffusion into some of many
small domains of regular motion within the critical structure.

A rough estimate for unknown function $B(\nu )$ can be obtained as follows.
The transition time is primarily determined by the width 
$\delta y\sim (\Delta K)^2$ of the chaotic layer around destroyed 
critical curve \cite{meiss,chilos,chirev} while the diffusion time through this
layer $t_0\sim 1/\Delta K$ \cite{chispan,chilos,ruffo,tolosa}. Noise
decreases this time down to $t_{\nu}\sim (\delta y)^2/\nu^2$. Hence,
the crossover corresponds to $t_{\nu}\sim t_0$,
whence:
\begin{equation}
B(\nu )\,\approx\,A\cdot\nu^{b} 
\label{noise}
\end{equation}
with $b=2/5$. Fitting the empirical data in Fig.2 in log--log scale gives:
$b\approx 0.39\pm 0.012$, which is surprisingly close to the theoretical 
estimate, and $A\approx 0.9716\pm 0.054$ (Fig.3).
The fitting accuracy is also fairly good: the relative $rms=0.019$.
Moreover, below crossover ($\Delta K < B(\nu )$) the width 
$\delta y$ as well as the diffusion time depend on $\nu$ only, 
and hence the transition time remains approximately
constant for a given $\nu$ (Fig.2). In any event, 
the minimal $(\Delta K)_{min}$
(\ref{abac}), which is the main result of the present study, is well above
the expected limitation for the double--precision
computation (see Fig.3).

In conclusion, the direct approach {\it a la} \cite{chirikov} to the problem
of the critical perturbation in the standard map does further confirm
Greene's hypothesis $K_{cr}=K_g$ with a much better exact upper bound
(\ref{abac}): $K_{cr}-K_g\,<\,2.5\times 10^{-4}$.

Still another recent confirmation of this conjecture (curiously, with
roughly the same statistical accuracy (\ref{stac})) has been inferred from
a detailed study of the critical structure at the chaos--chaos border in
standard map for $K=K_g$ \cite{tolosa}.

\bigskip

{\bf Acknowledgements.} I am grateful to D.L. Shepelyansky for interesting
discussions.
This work was partially supported by the Russia
Foundation for Fundamental Research, grant 97--01--00865.

\end{document}